\documentclass[journal]{IEEEtran}
\usepackage{cite}
\usepackage{amsmath,amssymb,amsfonts}
\usepackage{algorithmic}
\usepackage{graphicx} 
\usepackage{epsfig}
\usepackage{epstopdf}
\usepackage{textcomp}
\usepackage{xcolor}
\usepackage{algorithm}  
\usepackage{textcomp}
\usepackage{xcolor}
\usepackage{bm}
\usepackage{amsthm}
\usepackage{setspace}
\usepackage{stfloats}
\usepackage{caption}
\usepackage{subcaption}
\ifCLASSINFOpdf
 
\else
 
\fi

\begin{document}

\title{Cram\'{e}r-Rao Bound Minimization for Movable Antenna-Assisted Multiuser Integrated Sensing and Communications}

\author{Haoran~Qin,~Wen~Chen,~Qingqing~Wu,~Ziheng~Zhang,~Zhendong~Li,~Nan Cheng

\thanks{This work is supported by National key project 2020YFB1807700, NSFC 62071296, Shanghai Kewei 22JC1404000.(\emph{Corresponding author: Wen Chen.})

H. Qin, W. Chen, Q. Wu and Z. Zhang are with the Department of Electronic Engineering, Shanghai Jiao Tong University, Shanghai 200240, China (e-mail: haoranqin@sjtu.edu.cn; wenchen@sjtu.edu.cn; qingqingwu@sjtu.edu.cn; zhangziheng@sjtu.edu.cn).

Z. Li is with the School of Information and Communication Engineering, Xi'an Jiaotong University, Xi'an 710049, China (email: lizhendong@xjtu.edu.cn).

N. Cheng is with the State Key Lab. of ISN and School of Telecommunications
 Engineering, Xidian University, Xi’an 710071, China (e-mail:
dr.nan.cheng@ieee.org).

}}

\maketitle

\begin{abstract}
This paper investigates a movable antenna (MA)-assisted multiuser integrated sensing and communication (ISAC) system, where the base station (BS) and communication users are all equipped with MA for improving both the sensing and communication performance. We employ the Cramér-Rao bound (CRB) as the performance metric of sensing, thus a joint beamforming design and MAs' position optimizing problem is formulated to minimize the CRB. However the resulting optimization problem is NP-hard and the variables are highly coupled. To tackle this problem, we propose an alternating optimization (AO) framework by adopting semidefinite relaxation (SDR) and successive convex approximation (SCA) technique. Numerical results reveal that the proposed MA-assisted ISAC system achieves lower estimation CRB compared to the fixed-position antenna (FPA) counterpart. 

\end{abstract}

\begin{IEEEkeywords}
Movable antenna (MA), antenna positioning, Cramér-Rao bound, alternating optimization (AO).
\end{IEEEkeywords}

\IEEEpeerreviewmaketitle

\section{Introduction}
With the exponential growth of various sensing applications, such as gesture recognition, environment monitoring, and simultaneous localization and mapping, integrated sensing and communications (ISAC) technology is expected to play an increasingly important role in the next generation of wireless networks. By sharing hardware and spectrum resources to simultaneously provide ubiquitous sensing and communication services, ISAC is enable to greatly improve energy, spectral and hardware efficiency\cite{ziheng}.

Whether aiming for more accurate sensing capability or higher channel capacity for data transmissions, large-scale array antennas are often indispensable. However, their hardware cost and power consumption also increase proportionally with the number of radio frequency (RF) chains, which is contrary to the low-carbon and environmentally friendly requirements of the forthcoming sixth-generation (6G) communications. In addition, positions of traditional antenna array are fixed while the optimal array geometries for different sensing and communication applications may change, making it difficult to adapt to various business scenarios.

Recently, a novel antenna system, namely movable antenna (MA) or fluid antennas has gathered great interest due to its ability to take full advantage of signal variations in the spatial domain by flexibly adjusting its position\cite{FAS}. Extensive literature have investigated the significant benefits of MA in various communication systems, such as multiple input and multiple output (MIMO) communication system\cite{mimocapacity}, multicast communication\cite{gaoYing} and multiuser downlink communication\cite{qin2024antenna,wangHongHao}. In addition, \cite{maSensing} firstly introduced MA into wireless sensing system for improving sensing accuracy and explored the relationship between Cramér-Rao bound (CRB) and antenna positioning vector (APV). Nevertheless, MA-assisted multiuser ISAC system is still in its nascent stages.

In this paper, we design and optimize the MA-assited ISAC system, where the dual-functional radar-communication (DFRC) base station (BS) and user end are all equipped with MA. Firstly, we adopt CRB as sensing performance metrics and derive the function expression between CRB and antenna position. Then, we formulate a problem to minimize the CRB by jointly optimizing the transmit beamforming and the position of MA at different users and the BS, subject to the minimum signal-to-interference-plus-noise ratio (SINR) constraint, the limited moving region constraints and the total transmit power. To solve the resulting non-convex problem, an alternating optimisation (AO) algorithm based on successive convex approximation (SCA) and semidefinite relaxation (SDR) is proposed to obtain a sub-optimal solution. 

\section{System Model and Problem Formulation}
As shown in Fig. 1, a DFRC BS equipped with total $N_t$ transmit FPAs and $N_r$ receive MAs, serves $K$ users and detects one point-like target at the direction of $\theta $ simultaneously. The receive MAs at the BS can move freely within the given 1D line segment of length $L$. The APV of transmit antenna and receive antenna are denoted as 
${{\bf{d}}_t} = {\left[ {{d_{t,1}},{d_{t,2}}, \cdots ,{d_{t,N}}} \right]^T}$, ${{\bf{d}}_r} = {\left[ {{d_{r,1}},{d_{r,2}}, \cdots ,{d_{r,N}}} \right]^T}$ respectively. Meanwhile, each user is equipped with a single MA moving at the given local two-dimensional (2D) region ${{\cal C}_k}$, i.e., ${{\cal C}_k}=\left[ {x_k^{\min },x_k^{\max }} \right] \times [y_k^{\min } \times y_k^{\max }],1 \le k \le K$. 
\begin{figure}
 	\includegraphics[width=8.5cm, height=5cm]{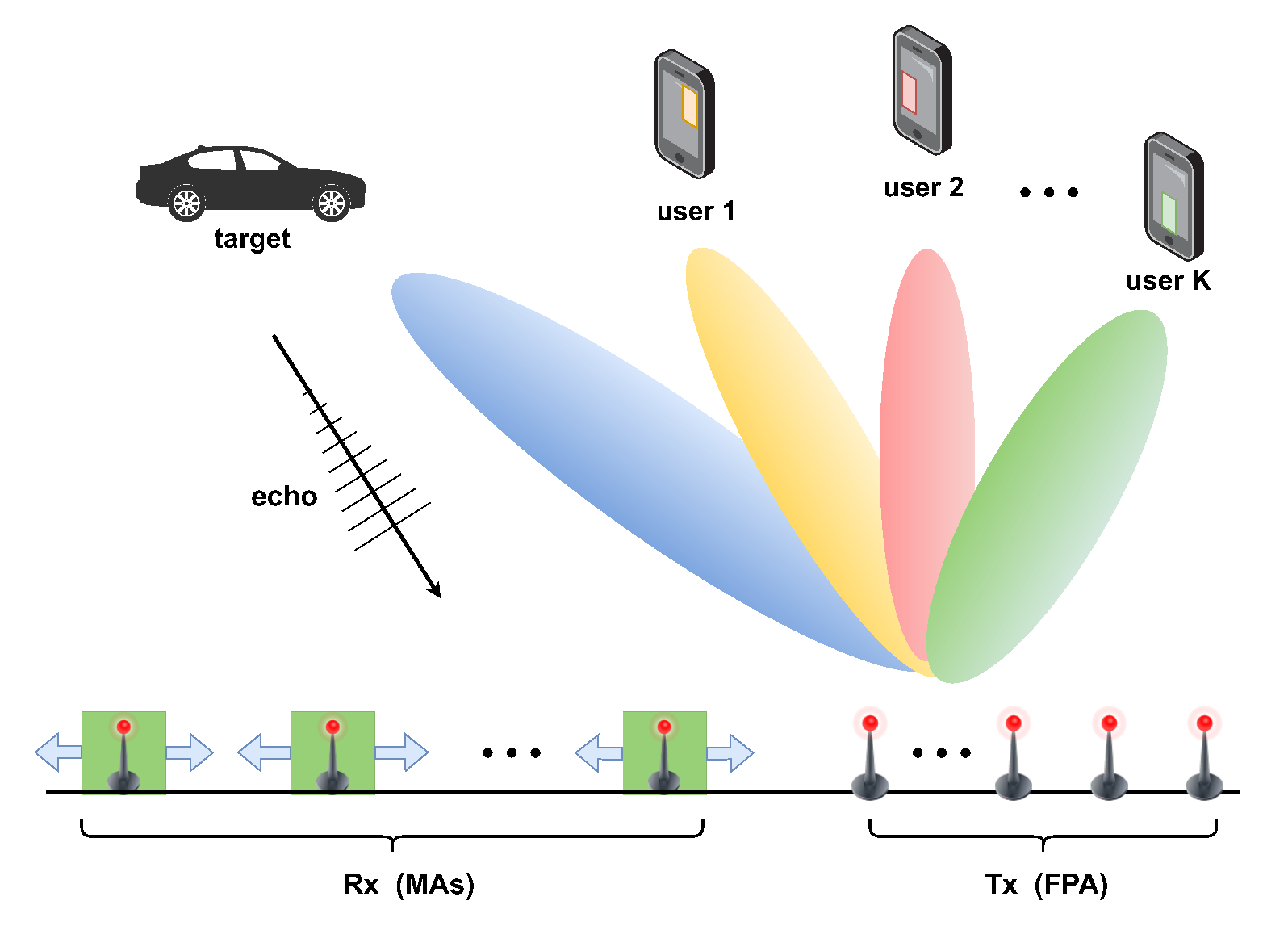}
	\caption{The MA-assisted multiuser ISAC system.}
	\label{Fig1}
\end{figure}
\subsection{Communication Signal Model}
Assuming that the channels between the BS and all users experience quasi-static flat-fading, the signal propagation phase difference of the $i$-th receive path between the $k$-user's MA and the origin of the receive region can be denoted as $\rho _{k,i}^r\left( {{\bf{u}}_k} \right) = {x_k}\sin \theta _{k,i}^r\cos \phi _{k,i}^r + {y_k}\cos \theta _{k,i}^r$. Similarly, the phase difference of the $j$-th transmit path between the $v$-th FPA and the origin of the transmit region is denoted as $\rho _{k,j}^t\left( {{\bf{v}}_n} \right) = {x_n}\sin \theta _{k,j}^t\cos \phi _{k,j}^t + {y_n}\cos \theta _{k,j}^t$ where $\theta _{k,i}^r$, $\phi _{k,i}^r$ and $\theta _{k,j}^t$, $\phi _{k,j}^t$ are denoted as elevation and azimuth AoAs/AoDs, respectively\cite{zhu2022modeling}. Thus, the channel vector from the BS to the $k$-th user can be denoted as
\begin{equation}
    \begin{aligned}
        {\bf{h}}_k^H({{\bf{u}}_k}) = {{\bf{f}}_k}{\left( {{{\bf{u}}_k}} \right)^H}{{\bf{\Sigma }}_k}{{\bf{T}}_k},
    \end{aligned}
\end{equation}
where ${{\bf{f}}_k}\left( {{{\bf{u}}_k}} \right) = {\left[ {{e^{j\frac{{2\pi }}{\lambda }\rho _{k,1}^r({{\bf{u}}_k})}},{e^{j\frac{{2\pi }}{\lambda }\rho _{k,2}^r({{\bf{u}}_k})}}, \ldots ,{e^{j\frac{{2\pi }}{\lambda }\rho _{k,Lr}^r({{\bf{u}}_k})}}} \right]^T}$ is the receive field response vector for the $k$-th user, ${{\bf{T}}_k} = \left[ {{{\bf{t}}_{k,1}},{{\bf{t}}_{k,2}}, \cdots ,{{\bf{t}}_{k,N}}} \right]$ represents the transmit field response matrix, and ${{\bf{t}}_{k,n}} = {\left[ {{e^{j\frac{{2\pi }}{\lambda }\rho _{k,1}^t({{\bf{v}}_n})}},{e^{j\frac{{2\pi }}{\lambda }\rho _{k,2}^t({{\bf{v}}_n})}}, \ldots ,{e^{j\frac{{2\pi }}{\lambda }\rho _{k,L_k^t}^t({{\bf{v}}_n})}}} \right]^T}$ is the transmit field response vector associated with the $n$-th transmit FPA, and $\lambda $ is the carrier wavelength. It's noted that ${{\bf{\Sigma }}_k}$ and ${{\bf{T}}_k}$ are both constant while ${{\bf{f}}_k}\left( {{{\bf{u}}_k}} \right)$ varies with user $k$'s antenna position\cite{zhu2022modeling}.
\newcounter{my1}
\normalsize
 \setcounter{my1}{\value{equation}}
 \setcounter{equation}{6}
  \begin{figure*}[!b]
  \hrulefill
  \vspace*{3pt}
    \begin{small}
        \begin{align}
            {\rm{CRB}}(\theta ) = \frac{{\sigma _R^2}}{{2{{\left| \alpha  \right|}^2}L\left( {{\rm{tr}}\left( {{\bf{AD}}_t^T{{\bf{R}}_X}{\rm{tr}}({\bf{D}}_r^H) + {\bf{A}}{{\bf{R}}_X}{\rm{tr}}({\bf{D}}_r^H{{\bf{D}}_r}) + {{\bf{D}}_t}{\bf{AD}}_t^H{{\bf{R}}_X}{N_r} + {\rm{tr}}({{\bf{D}}_r}){{\bf{D}}_t}{\bf{A}}{{\bf{R}}_X}} \right) - \frac{{{{\left| {{\rm{tr}}\left( {\left( {{\rm{tr}}({\bf{D}}_r^H){\bf{A}} + {N_r}{{\bf{D}}_t}{\bf{A}}} \right){{\bf{R}}_X}} \right)} \right|}^2}}}{{{\rm{tr}}\left( {{N_r}{\bf{A}}{{\bf{R}}_X}} \right)}}} \right)}}
     \end{align}
    \end{small}
    \end{figure*}
 \setcounter{equation}{\value{my1}}

Let ${\bf{W}} = \left[ {{{\bf{w}}_1},{{\bf{w}}_2}, \cdots ,{{\bf{w}}_K}} \right]$ denote the beamforming matrix with the $k$-th column being the beamformer for the $k$-th user. The transmitted signal by the BS at time slot $t$ can be denoted as ${\bf{x}}(t) = {\bf{Ws}}(t)$, where ${\bf{s}}(t)$ contains $K$ unit-power data streams intended for the $K$ users. The additive white Gaussian noise (AWGN) at the $k$-th user with an average noise power of ${\sigma ^2}$ is given by ${n_k} \sim {\cal C}{\cal N}(0,{\sigma ^2})$, thus we have the received signal of the user $k$ at time slot $t$ as
\begin{equation}
    \begin{aligned}
        {{\bf{y}}_k}(t) = {{\bf{h}}_k}{({{\bf{u}}_k})^H}{\bf{Ws}}(t) + {n_k}(t),
    \end{aligned}
\end{equation}
and its SINR is given as 
\begin{small}
\begin{equation}
    \begin{aligned}
        {\gamma _k} = \frac{{{{\left| {{{\mathbf{h}}_k}{{\left( {{{\mathbf{u}}_k}} \right)}^H}{{\mathbf{w}}_k}} \right|}^2}}}{{\sum\limits_{q = 1,q \ne k}^K {\left| {{{\mathbf{h}}_k}{{\left( {{{\mathbf{u}}_k}} \right)}^H}{{\mathbf{w}}_q}} \right| + {\sigma ^2}} }},\forall k.
    \end{aligned}
\end{equation}
\end{small}

\subsection{Radar Signal Model}
In this case, we consider a scenario where the target is modeled as an unstructured point that is far away from the BS. The target response matrix can be written as ${\bf{G}}(\theta ) = \alpha {\bf{b}}(\theta ){{\bf{a}}^H}(\theta )$, where ${\bf{a}}\left( \theta  \right) = {\left[ {{e^{j\frac{{2\pi }}{\lambda }{d_{t,1}}\sin \theta }},{e^{j\frac{{2\pi }}{\lambda }{d_{t,2}}\sin \theta }}, \cdots ,{e^{j\frac{{2\pi }}{\lambda }{d_{t,N}}\sin \theta }}} \right]^T}$, ${\bf{b}}\left( \theta  \right) = {\left[ {{e^{j\frac{{2\pi }}{\lambda }{d_{r,1}}\sin \theta }},{e^{j\frac{{2\pi }}{\lambda }{d_{r,2}}\sin \theta }}, \cdots ,{e^{j\frac{{2\pi }}{\lambda }{d_{r,N}}\sin \theta }}} \right]^T}$ denote the transmit steering vector and receive steering vector, respectively, and $\alpha$ represents the reflection coefficient. By stacking the  transmit symbols, the received  echo signals, and the noise over the radar dwell time, we can have 
\begin{small}
\begin{equation}
    \begin{aligned}
        {\bf{Y}} = {\bf{G}}(\theta ){\bf{X}} + {\bf{N}},
    \end{aligned}
\end{equation}
\end{small}where  ${\bf{X}} = \left[ {{\bf{x}}(1),{\bf{x}}(2), \cdots ,{\bf{x}}(L)} \right]$, ${\bf{Y}}$ and ${\bf{N}}$ have the similar form. ${\bf{N}}$ denotes the noise matrix at the BS receive antennas with variance of each entry being ${\sigma _R^2}$. Assuming the data streams are independent of each other and $L$ to be sufficiently large, the sample covariance matrix of the transmitted signal can be approximated as
\begin{equation}
    \begin{aligned}
        {{\bf{R}}_X} = \frac{1}{L}\sum\limits_{t = 1}^L {{\bf{x}}(t){{\bf{x}}^H}(t)}  \approx {\bf{W}}{{\bf{W}}^H}.
    \end{aligned}
\end{equation}

According to\cite{liu2021cramer}, the CRB for estimating the DoA $\theta $ of the point target is given as follow
\begin{equation}
    \begin{aligned}
        {\rm{CRB}}(\theta ) = \frac{{\sigma _R^2}}{{2{{\left| \alpha  \right|}^2}L\left( {{\rm{tr}}({{{\bf{\dot G}}}^H}(\theta ){\bf{\dot G}}(\theta ){{\bf{R}}_{\bf{X}}}) - \frac{{{{\left| {{\rm{tr}}({{{\bf{\dot G}}}^H}(\theta ){\bf{G}}(\theta ){{\bf{R}}_{\bf{X}}})} \right|}^2}}}{{{\rm{tr}}({{\bf{G}}^H}(\theta ){\bf{G}}(\theta ){{\bf{R}}_{\bf{X}}})}}} \right)}}.      
    \end{aligned}
\end{equation}
To gain more insight and facilitate the antenna position and beamforming design, we re-express ${\rm{CRB}}(\theta )$ in (6) into a more intuitive form w.r.t. the APV ${{\bf{d}}_r}$. Toward this end, we have
\begin{equation}
    \begin{aligned}
        {{\bf{G}}^H}(\theta ){\bf{G}}(\theta ){{\bf{R}}_X}= {\left| \alpha  \right|^2}{\bf{a}}(\theta ){{\bf{b}}^H}(\theta ){\bf{b}}(\theta ){{\bf{a}}^H}(\theta ){{\bf{R}}_X}= {\left| \alpha  \right|^2}{N_r}{\bf{A}}{{\bf{R}}_X}, \nonumber
    \end{aligned}
\end{equation}
\begin{equation}
    \begin{aligned}
     {{{\bf{\dot G}}}^H}(\theta ){\bf{G}}(\theta ){{\bf{R}}_X} &= {\left| \alpha  \right|^2}\left( {{\bf{a}}(\theta ){{{\bf{\dot b}}}^H}(\theta ) + {\bf{\dot a}}(\theta ){{\bf{b}}^H}(\theta )} \right){\bf{b}}(\theta ){{\bf{a}}^H}(\theta ){{\bf{R}}_X}\nonumber\\
{\rm{~~}} &= {\left| \alpha  \right|^2}\left( {{\rm{tr}}({\bf{D}}_r^H){\bf{A}} + {N_r}{{\bf{D}}_t}{\bf{A}}} \right){{\bf{R}}_X},      
    \end{aligned}
\end{equation}
\begin{equation}
    \begin{aligned}
    &{{{\bf{\dot G}}}^H}(\theta ){\bf{\dot G}}(\theta ){{\bf{R}}_X}= {\left| \alpha  \right|^2}\left( {{\bf{a}}(\theta ){{{\bf{\dot b}}}^H}(\theta ) + {\bf{\dot a}}(\theta ){{\bf{b}}^H}(\theta )} \right) \nonumber\\
    &\quad \left( {{\bf{b}}(\theta ){{{\bf{\dot a}}}^H}(\theta ) + {\bf{\dot b}}(\theta ){{\bf{a}}^H}(\theta )} \right){{\bf{R}}_X} \\
    &\quad = {\left| \alpha  \right|^2}\Big( {\bf{AD}}_t^T{{\bf{R}}_X}{\rm{tr}}({\bf{D}}_r^H) + {\bf{A}}{{\bf{R}}_X}{\rm{tr}}({\bf{D}}_r^H{{\bf{D}}_r})  \\
    &\quad\quad +{{\bf{D}}_t}{\bf{AD}}_t^H{{\bf{R}}_X}{N_r} + {\rm{tr}}({{\bf{D}}_r}){{\bf{D}}_t}{\bf{A}}{{\bf{R}}_X} \Big),
    \end{aligned}
\end{equation}
where ${{\bf{D}}_t} = j\frac{{2\pi }}{\lambda }{\rm{diag}}({{\bf{d}}_t})\cos \theta $, ${{\bf{D}}_r} = j\frac{{2\pi }}{\lambda }{\rm{diag}}({{\bf{d}}_r})\cos \theta $, ${\bf{A}} = {\bf{a}}(\theta ){{\bf{a}}^H}(\theta )$. Therefore, the ${\rm{CRB}}(\theta )$ can be rewritten as (7) shown in the bottom at this page.
\subsection{Problem Formulation}
In this paper, we aim to minimize the ${\rm{CRB}}(\theta)$ by jointly optimizing the transmit beamforming matrix and the position of antenna at the user and the BS. Thus the corresponding optimization problem can be expressed as
\setcounter{equation}{7}
\begin{small}
\begin{subequations}
      \begin{align} 
 \left( {{\textrm{P0}}} \right){\rm{~~}} \mathop {\min }\limits_{{\bf{W}},{{\bf{d}}_r},\left\{ {{{\bf{u}}_k}} \right\}_{k = 1}^K}{\rm{~~}}&{\rm{CRB}}(\theta )\nonumber\\ 
{\rm{s.t.}}\qquad&\left\| {\bf{W}} \right\|_F^2 \le {P_T},\\
\quad&{\gamma _k} \ge {\gamma _{th}},\forall k,\\
\quad&{{\bf{u}}_k} \in {{\cal C}_k},\forall k,\\
\quad&{{{d}}_{r,1}}{\rm{ }} \ge {\rm{0}}, {{{d}}_{r,Nr}} \le {d_{\max }},\\
{{{d}}_{r,n}}&- {{{d}}_{r,n - 1}}{\rm{ > }}{d_{\min }},n = 2,3, \cdots ,{N_r},
\end{align}
\end{subequations}
\end{small}in which constraint (8a) is the limited transmit power constraint and (8b) represents the minimum SINR requirement of each user. (8c) is the moving region constraint for each user's MA. (8d) ensures the receive MA at BS are moving within a 1D line segment $\left[ {0,{d_{\max }}} \right]$ and condition (8e) guarantees a minimum distance between adjacent antennas to avoid antenna coupling. Evidently, given the non-convex nature of both constraint (8b) and the objective function, this problem is inherently non-convex and poses a challenge for direct solution.

\section{Joint Antenna Position and Beamforming Design For CRB Minimization}
In this section, we jointly optimize the antenna position and transmit beamforming matrix to minimize the CRB for estimating the DoA of the target. Specifically, we decouple the orginal problem (P0) into three sub-problems and iteratively optimize transmit beamformer ${\bf{W}}$, the APV of BS receive antenna ${{{\bf{d}}_r}}$ and users' antenna position $\left\{ {{{\bf{u}}_k}} \right\}_{k = 1}^K$ in an alternate manner with the other being fixed until the objective function converges.

\subsection{Transmit Beamforming Optimization}
First, we optimize the transmit beamformer ${\bf{W}}$ in problem (P0) under the assumption that the positions of all antennas are fixed. In this case, the transmit beamforming optimization problem is formulated as
\begin{subequations}
  \begin{align}
  \left( {{\textrm{P1}}} \right){\rm{~~}}\mathop {\max }\limits_{\bf{W}} &{\rm{~~tr}}({{\bf{D}}_t}{\bf{A}}{{\bf{R}}_X}){\rm{tr}}({{\bf{D}}_r}) + {\rm{tr}}({\bf{A}}{{\bf{R}}_X}){\rm{tr}}({\bf{D}}_r^H{{\bf{D}}_r}) \nonumber\\
  &+ {\rm{tr}}({\bf{AD}}_t^H{{\bf{R}}_X}){\rm{tr}}({\bf{D}}_r^H)\nonumber\\
{\rm{ }}& - \frac{{{{\left| {{N_r}{\rm{tr}}({{\bf{D}}_t}{\bf{A}}{{\bf{R}}_X}) + {\rm{tr}}({\bf{D}}_r^H){\rm{tr}}({\bf{A}}{{\bf{R}}_X})} \right|}^2}}}{{{N_r}{\rm{tr}}({\bf{A}}{{\bf{R}}_X})}}\nonumber\\
{\rm{s.t.}}{\rm{~~}}&{\gamma _k} \ge {\gamma _{th}},\forall k,\\
{\rm{}}&\sum\limits_{k = 1}^K {{\rm{tr}}\left( {{{\bf{w}}_k}{\bf{w}}_k^H} \right)}  \le {P_T},\\
{\rm{}}&{{\bf{R}}_X} = \sum\limits_{k = 1}^K {{{\bf{w}}_k}{\bf{w}}_k^H}.
  \end{align}
\end{subequations} 
 By taking a closer look, we see that the objective function is non-convex in ${\bf{W}}$ due to its fractional structure. Fortunately, we can transform it into a convex expression with respect to ${{\bf{R}}_X}$ by applying the Schur complement condition. For simplicity of expression, we let
${{\bf{M}}_{11}} = {\rm{tr}}({{\bf{D}}_t}{\bf{A}}{{\bf{R}}_X}){\rm{tr}}({{\bf{D}}_r}) + {\rm{tr}}({{\bf{D}}_t}{\bf{AD}}_t^H{{\bf{R}}_X}){N_r} + {\rm{tr}}({\bf{A}}{{\bf{R}}_X}){\rm{tr}}({\bf{D}}_r^H{{\bf{D}}_r}) + {\rm{tr}}({\bf{AD}}_t^H{{\bf{R}}_X}){\rm{tr}}({\bf{D}}_r^H),
{{\bf{M}}_{12}} = {N_r}{\rm{tr}}({{\bf{D}}_t}{\bf{A}}{{\bf{R}}_X}) + {\rm{tr}}({\bf{D}}_r^H){\rm{tr}}({\bf{A}}{{\bf{R}}_X}),
{{\bf{M}}_{21}} = {N_r}{\rm{tr}}({\bf{D}}_t^H{\bf{A}}{{\bf{R}}_X}) + {\rm{tr}}({{\bf{D}}_r}){\rm{tr}}({\bf{A}}{{\bf{R}}_X}),
{{\bf{M}}_{22}} = {N_r}{\rm{tr}}({\bf{A}}{{\bf{R}}_X}),
{\rm{}}{{\bf{R}}_X} = \sum\limits_{k = 1}^K {{{\bf{W}}_k}},
{{\bf{W}}_k} = {{\bf{w}}_k}{\bf{w}}_k^H.$
By introducing an auxiliary variable $t$, problem (P1) is equivalently re-expressed as
\begin{subequations}
    \begin{align}
 \left( {{\textrm{P1.1}}} \right){\rm{~~}}\mathop {\min }\limits_{\left\{ {{{\bf{w}}_k}} \right\}_{k = 1}^K,{{\bf{R}}_X},t} {\rm{~}} &- t\nonumber\\
{\rm{s.t.}}{\rm{~~~~~}}&\left[ {\begin{array}{*{20}{c}}
{{{\bf{M}}_{11}} - t}&{{{\bf{M}}_{12}}}\\
{{{\bf{M}}_{21}}}&{{{\bf{M}}_{22}}}
\end{array}} \right] \succeq {\bf{0}},\\
&{\rm{(9a)}} \sim {\rm{(9c)}}.
    \end{align}
\end{subequations}
While problem (P1.1) remains non-convex, it can be relaxed into a convex problem using the classical Semidefinite Relaxation (SDR) technique. However, the optimal solution requires ${\rm{rank}}({{\bf{W}}_k}) = 1$, and ${{\bf{W}}_k} \succeq 0$. The relaxation of this rank one constraint allows for the resolution of problem (P1.1) in the form of a semi-definite program (SDP) as problem (P1.2), which can be optimally solved by means of numerical tools such as CVX.
\begin{subequations}
  \begin{align}
 \left( {{\textrm{P1.2}}} \right){\rm{~~}}\mathop {\min }\limits_{\left\{ {{{\bf{W}}_k}} \right\}_{k = 1}^K,{t}} {\rm{}}& - {t}\nonumber\\
{\rm{s.t.}}{\rm{~~~}}&\left[ {\begin{array}{*{20}{c}}
{{{\bf{M}}_{11}} - {t}}&{{{\bf{M}}_{12}}}\\
{{{\bf{M}}_{21}}}&{{{\bf{M}}_{22}}}
\end{array}} \right] \succ {\bf{0}},\\
&{\rm{ tr(}}{{\bf{W}}_k}{{\bf{H}}_k}{\rm{)}} \ge\nonumber\\
{\gamma _k}{\sigma ^2} +& {\gamma _k}\left( {\sum\limits_{q = 1,q \ne k}^K {{\rm{tr(}}{{\bf{W}}_q}{{\bf{H}}_k}{\rm{)}}} } \right),\forall k,\\
{\rm{~}}&\sum\limits_{k = 1}^K {{\rm{tr}}\left( {{{\bf{W}}_k}} \right)}  \le {P_T}.
   \end{align}
\end{subequations} 

\subsection{The Position of BS Antenna Optimization}
Next, we optimize the position of receive antenna at BS under any given transmit beamformer ${\bf{W}}$ and the fixed position of antenna at user $\left\{ {{{\bf{u}}_k}} \right\}_{k = 1}^K$. In this case, the optimization problem w.r.t APV ${{\bf{d}}_{r}}$ can be expressed as
\begin{subequations}
\begin{align}
 \left( {{\textrm{P2}}} \right){\rm{~~}}\mathop {\max }\limits_{{{\bf{d}}_r}} {\rm{~~}}&{\rm{tr}}({{\bf{D}}_t}{\bf{A}}{{\bf{R}}_X}){\rm{tr}}({{\bf{D}}_r}) + {\rm{tr}}({\bf{A}}{{\bf{R}}_X}){\rm{tr}}({\bf{D}}_r^H{{\bf{D}}_r})\nonumber\\
&+ {\rm{tr}}({\bf{AD}}_t^H{{\bf{R}}_X}){\rm{tr}}({\bf{D}}_r^H)\nonumber\\
{\rm{}} &- \frac{{{{\left| {{N_r}{\rm{tr}}({{\bf{D}}_t}{\bf{A}}{{\bf{R}}_X}) + {\rm{tr}}({\bf{D}}_r^H)\rm{tr}({\bf{A}}{{\bf{R}}_X})} \right|}^2}}}{{{N_r}{\rm{tr}}({\bf{A}}{{\bf{R}}_X})}}\nonumber\\
{\rm{s.t.}}{\rm{~~}}&{{\bf{d}}_{r,1}}{\rm{ }} \ge {\rm{0,     }}{{\bf{d}}_{r,Nr}} \le {d_{\max }},\\
{\rm{~~}}&{{\bf{d}}_{r,n}} - {{\bf{d}}_{r,n - 1}}{\rm{ > }}{d_{\min }},n = 2,3, \cdots ,{N_r}.
\end{align}
\end{subequations}
To have a clearer intuition, we rewrite the objective function as a function w.r.t. ${{\bf{d}}_{r}}$ as following
\begin{small}
\begin{equation}
    \begin{aligned}
    &{\rm{tr}}({{\bf{D}}_t}{\bf{A}}{{\bf{R}}_X}){\rm{tr}}({{\bf{D}}_r}) =  - \frac{{4{\pi ^2}}}{{{\lambda ^2}}}{\cos ^2}\theta {\bf{d}}_t^H({\bf{A}}{{\bf{R}}_X} \odot {\bf{I}}){{\mathbf{J}}_{{{N}_{t}},{{N}_{r}}}}{{\bf{d}}_r},\\
    &{\rm{tr}}({\bf{AD}}_t^H{{\bf{R}}_X}){\rm{tr}}({\bf{D}}_r^H)=-\frac{4{{\pi }^{2}}}{{{\lambda }^{2}}}{{\cos }^{2}}\theta \mathbf{d}_{t}^{H}\text{(}\mathbf{A}{{\mathbf{R}}_{X}}\odot \mathbf{I}\text{)}{{\mathbf{J}}_{{{N}_{t}},{{N}_{r}}}}{{\mathbf{d}}_{r}},\\
    &{\left| {{N_r}tr({{\bf{D}}_t}{\bf{A}}{{\bf{R}}_X}) + tr({\bf{D}}_r^H)tr({\bf{A}}{{\bf{R}}_X})} \right|^2} = {\bf{d}}_r^H{\bf{E}}{{\bf{d}}_r} + {{\bf{f}}^T}{{\bf{d}}_r}\\
    &+ {\left| {{N_r}tr({{\bf{D}}_t}{\bf{A}}{{\bf{R}}_X})} \right|^2},
    {\rm{~~}}{\bf{E}} = \frac{{4{\pi ^2}}}{{{\lambda ^2}}}{\cos ^2}\theta {\rm{t}}{{\rm{r}}^2}({\bf{A}}{{\bf{R}}_X}){{\bf{J}}_{{N_r},{N_r}}},\\
    &{\bf{f}} = - \frac{{8{\pi ^2}}}{{{\lambda ^2}}}{\cos ^2}\theta {\rm{tr}}({\bf{A}}{{\bf{R}}_X}){N_r}{\mathop{\rm Re}\nolimits} \{ {\rm{tr}}({\rm{diag}}({{\bf{d}}_t}){\bf{A}}{{\bf{R}}_X})\} {{\bf{1}}_{{N_r}}},\nonumber
    \end{aligned}
\end{equation}
\end{small}where ${{\mathbf{J}}_{{{N}_{t}},{{N}_{r}}}}$ represents the ${N_t} \times {N_r}$ dimensional all-one matrix, ${{\bf{1}}_{{N_r}}}$ denote the ${N_t} \times 1$ dimensional all-one vector. The matrix ${\bf{E}}$ is semidefinite because it has only one positive eigenvalue, while the rest of the eigenvalues are all zero. It's noted that the objective function is not concave due to the non-concavity of the second term ${\rm{tr}}({\bf{A}}{{\bf{R}}_X}){\rm{tr}}({\bf{D}}_r^H{{\bf{D}}_r})$. To address this issue, successive convex approximation (SCA) method is adopted to obtain a lower bound for the objective function. Let ${{\bf{d}}_r^{(i)}}$ denote the APV of the MAs at BS in the $i$-th iteration of SCA, we have the following inequality: ${\rm{tr}}({\bf{A}}{{\bf{R}}_X}){\rm{tr}}({\bf{D}}_r^H{{\bf{D}}_r}) = \frac{{4{\pi ^2}}}{{{\lambda ^2}}}{\cos ^2}\theta {\bf{d}}_r^H{\rm{tr}}({\bf{A}}{{\bf{R}}_X}){{\bf{I}}_r}{{\bf{d}}_r}
 \ge \frac{{4{\pi ^2}}}{{{\lambda ^2}}}{\cos ^2}\theta {\rm{tr}}({\bf{A}}{{\bf{R}}_X})\left( {{\bf{d}}_r^{(i)H}{\bf{d}}_r^{(i)} + 2{\bf{d}}_r^{(i)H}({{\bf{d}}_r} - {\bf{d}}_r^{(i)})} \right)$.
 Therefore, this subproblem can be written as the following
\begin{subequations}
\begin{align}
    \left( {{\textrm{P2.1}}} \right)&\mathop {\max }\limits_{{{\bf{d}}_r}}{\rm{~}}-\frac{{4{\pi ^2}}}{{{\lambda ^2}}}{\cos ^2}\theta {\bf{d}}_t^H{\rm{(}}({{\bf{R}}_X}{\bf{A}} + {\bf{A}}{{\bf{R}}_X}) \odot {\bf{I}}{\rm{)}}{{\bf{J}}_{{N_t},{N_r}}}{{\bf{d}}_r} \nonumber\\
    &+ \frac{{4{\pi ^2}}}{{{\lambda ^2}}}{\cos ^2}\theta {\rm{tr}}({\bf{A}}{{\bf{R}}_X})\left( {{\bf{d}}_r^{(i)H}{\bf{d}}_r^{(i)} + 2{\bf{d}}_r^{(i)H}({{\bf{d}}_r} - {\bf{d}}_r^{(i)})} \right) \nonumber\\
   &-\frac{{{\bf{d}}_r^H{\bf{E}}{{\bf{d}}_r} + {{\bf{f}}^T}{{\bf{d}}_r} + {{\left| {{N_r}{\rm{tr}}({{\bf{D}}_t}{\bf{A}}{{\bf{R}}_X})} \right|}^2}}}{{{N_r}{\rm{tr}}({\bf{A}}{{\bf{R}}_X})}}\nonumber\\
&{\rm{s.t.}}{\rm{~~}}{{\bf{d}}_{r,1}}{\rm{ }} \ge 0{\rm{,~~~}}{{\bf{d}}_{r,Nr}} \le {d_{\max }},\\
&{\rm{~~~~~~}}{{\bf{d}}_{r,n}} - {{\bf{d}}_{r,n - 1}}{\rm{ > }}{d_{\min }},n = 2,3, \cdots ,{N_r}.
\end{align}   
\end{subequations}
It's not difficult to observe that the resulting problem (P2.1) is concave w.r.t ${{{\bf{d}}_r}}$ and can be solved by optimization toolbox like CVX.

\subsection{The Position of User Antenna Optimization}
Thirdly, for given beamforming matrix ${{\bf{W}}}$ and fixed APV ${{\bf{d}}_r}$ at BS, the optimization problem for each user's antenna position can be implemented in parallel because the MAs position variables $\left\{ {{{\bf{u}}_k}} \right\}_{k = 1}^K$ are mutually independent. Thus, the sub-problems associated with user $k$ is formulated as
\begin{small}
\begin{subequations}
\begin{align}
\left( {{\textrm{P3}}} \right)
{\rm{~~}}{\rm{find~~}}&{{\bf{u}}_k}\nonumber\\
{\rm{s.t.}}{\rm{~~}}&{\left| {{{\bf{h}}_k}{{\left( {{{\bf{u}}_k}} \right)}^H}{{\bf{w}}_k}} \right|^2} \ge\nonumber\\
&{\gamma _k}\left( {\sum\limits_{q = 1,q \ne k}^K {\left| {{{\bf{h}}_k}{{\left( {{{\bf{u}}_k}} \right)}^H}{{\bf{w}}_q}} \right| + {\sigma ^2}} } \right),\\
{\rm{~~}}&{{\bf{u}}_k} \in {{\cal C}_k}.
    \end{align}
\end{subequations}
\end{small}  
To handle the non-convexity of the constraint (14a), the left hand side (LHS) and the right hand side (RHS) of this inequality are appropriately deflated by applying the SCA method. We firstly rewrite the RHS of the inequality as a display expression in terms of ${{\bf{u}}_k}$ as following
\begin{small}
\begin{equation}
    \begin{aligned}
      &{\varsigma _q}({{\bf{u}}_k}) = {\left| {{{\bf{h}}_k}{{\left( {{{\bf{u}}_k}} \right)}^H}{{\bf{w}}_q}} \right|^2}
 = {{\bf{f}}_k}{\left( {{{\bf{u}}_k}} \right)^H}{{\bf{\Sigma }}_k}{{\bf{T}}_k}{{\bf{W}}_q}{\bf{T}}_k^H{\bf{\Sigma }}_k^H{{\bf{f}}_k}\left( {{{\bf{u}}_k}} \right)\\
 &={\rm{tr}}({{\bf{A}}_{k,q}}) + 2\sum\limits_{i = 1}^{{L_r} - 1} {\sum\limits_{j = i + 1}^{{L_r}} {\left| {{{\bf{A}}_{k,q}}\left( {i,j} \right)} \right|\cos \left( {\psi (i,j,q)} \right)} },\nonumber
    \end{aligned}
\end{equation}   
\end{small}where $\psi (i,j,q) = \frac{{2\pi }}{\lambda }\left( {\rho _{k,i}^r\left( {{{\bf{u}}_k}} \right) - \rho _{k,j}^r\left( {{{\bf{u}}_k}} \right)} \right) + \angle {{\bf{A}}_{k,q}}(i,j)$, ${{\bf{A}}_{k,q}} \buildrel \Delta \over = {{\bf{\Sigma }}_k}{{\bf{T}}_k}{{\bf{W}}_q}{\bf{T}}_k^H{\bf{\Sigma }}_k^H$. We can re-express the LHS term ${\rm{ }}{\left| {{{\bf{h}}_k}{{\left( {{{\bf{u}}_k}} \right)}^H}{{\bf{w}}_k}} \right|^2}$ in a similar way.
With given local point ${{\bf{u}}_k^i}$ in the $i$-th iteration of SCA, a upper bound of the RHS is obtained by applying the Taylor's second order expansion theorem as
\begin{equation}
    \begin{aligned}
    \varsigma _q^{ub}({{\bf{u}}_k})&=\frac{{{\tau _{k,q}}}}{2}{({{\bf{u}}_k} - {\bf{u}}_k^i)^T}({{\bf{u}}_k} - {\bf{u}}_k^i),\\
    &+ \nabla {\varsigma _q}{({\bf{u}}_k^i)^T}({{\bf{u}}_k} - {\bf{u}}_k^i) + {\varsigma _q}({\bf{u}}_k^i),
    \end{aligned}
\end{equation} 
where $\nabla {\varsigma _q}({\bf{u}}_k^i) = {\left[ {{{\left. {\frac{{\partial {\varsigma _q}({{\bf{u}}_k})}}{{\partial {x_k}}}} \right|}_{{{\bf{u}}_k} = {\bf{u}}_k^i}},{{\left. {\frac{{\partial {\varsigma _q}({{\bf{u}}_k})}}{{\partial {y_k}}}} \right|}_{{{\bf{u}}_k} = {\bf{u}}_k^i}}} \right]^T}$. ${\tau _{k,q}} = \frac{{16{\pi ^2}}}{{{\lambda ^2}}}\sum\limits_{i = 1}^{{L_r} - 1} {\sum\limits_{j = i + 1}^{{L_r}} {\left| {{{\bf{A}}_{k,q}}\left( {i,j} \right)} \right|} }$ is determined from the frobenius norm of the Hessian matrix of ${\varsigma _q}({{\bf{u}}_k})$, as referenced in \cite{gaoYing,mimocapacity}.
By the same token, it's not difficult to find a lower bound of the LHS as following:
\begin{small}
\begin{equation}
    \begin{aligned}
    {\eta ^{lb}}({{\bf{u}}_k})=&-\frac{{{\delta _k}}}{2}{({{\bf{u}}_k} - {\bf{u}}_k^i)^T}({{\bf{u}}_k} - {\bf{u}}_k^i)\\
    &+ \nabla \eta {({\bf{u}}_k^i)^T}({{\bf{u}}_k} - {\bf{u}}_k^i) + \eta ({\bf{u}}_k^i),
    \end{aligned}
\end{equation}
\end{small}
where $\nabla \eta ({\bf{u}}_k^i) = {\left[ {{{\left. {\frac{{\partial \eta ({{\bf{u}}_k})}}{{\partial {x_k}}}} \right|}_{{{\bf{u}}_k} = {\bf{u}}_k^i}},{{\left. {\frac{{\partial \eta ({{\bf{u}}_k})}}{{\partial {y_k}}}} \right|}_{{{\bf{u}}_k} = {\bf{u}}_k^i}}} \right]^T}$, ${\delta _k} = \frac{{16{\pi ^2}}}{{{\lambda ^2}}}\sum\limits_{i = 1}^{{L_r} - 1} {\sum\limits_{j = i + 1}^{{L_r}} {\left| {{{\bf{A}}_{k,k}}(i,j)} \right|} }.$
In this case, the original problem can be relaxed to a convex problem as follow:
\begin{small}
\begin{subequations}
    \begin{align}
\left( {{\textrm{P3.1}}} \right)\quad
{\rm{find~~}}&{{\bf{u}}_k}\nonumber\\
{\rm{s.t.}}{\rm{~~}}{\eta ^{lb}}({{\bf{u}}_k}) &\ge {\gamma _{th}}\sum\limits_{q = 1,q \ne k}^K {\varsigma _q^{ub}({{\bf{u}}_k})}  + {\gamma _{th}}{\sigma ^2},\\
{\rm{~~}}&{{\bf{u}}_k} \in {{\cal C}_k}.
    \end{align}
\end{subequations}    
\end{small}  
Fortunately, it's a classical find problem with one linear constraint and one quadratic constraint which can be optimally solved using CVX.

Based on the above derivations, the overall algorithm for minimizing the CRB
is summarized in $\textbf{Algorithm 1}$.
\begin{small}
\begin{algorithm}[H]
		\caption{Joint antenna position and beamforming design algorithm}
		\begin{algorithmic}[1]
			\REPEAT
			\STATE Update the transmit beamfoming matrix  ${\bf{W}}$ by solving (P1.2).
			\STATE Update the antenna position ${{\bf{d}}_{r}}$ at BS by solving (P2.1).
			\STATE Update the antenna position $\left\{ {{{\bf{u}}_k}} \right\}_{k = 1}^K$ at each user by solving (P3.1).
			\UNTIL The fractional decrease of the objective value is below a threshold ${\varepsilon _1}$.
		\end{algorithmic}
\end{algorithm}
\end{small}
The complexity of each iteration for updating ${\bf{W}}$,  ${{\bf{d}}_{r}}$, $\left\{ {{{\bf{u}}_k}} \right\}_{k = 1}^K$ are given by ${\cal O}\left( {{K^{1.5}}N_t^{3.5}} \right)$, ${\cal O}\left( {{I_1}N_r^{1.5}} \right)$, ${\cal O}\left( {{K^{3.5}}} \right)$ respectively, where $I_1$ is the maximum numbers of iteration. Hence, the overall computational complexity of $\textbf{Algorithm  1}$ is ${\cal O}\left( {{I_2}\left( {{K^{1.5}}N_t^{3.5} + {I_1}N_r^{1.5} + {K^{3.5}}} \right)} \right)$, where $I_2$ represents the number of iterations required for convergence{\cite{wangHongHao}}.

\section{Simulation Results}
In the section, we provide numerical results to validate the performance of our proposed algorithm. We consider a DFRC BS that is equipped with $N_t = 10$ FPAs for transmission and $N_r = 10$ MAs for reception. The BS serves $K=5$ users and detects one target located 30 meters away in the direction of $\frac{\pi }{3}$. Assuming that the users are uniformly distributed within a range of $\left[ {20,60} \right]$ meters from the BS and each user have $L_k^t=10$ transmit paths and $L_k^r=10$ receive paths. The elevation and azimuth angles of AoAs/AoDs are modeled as independent and identically distributed (i.i.d.) variables, uniformly distributed over $\left[ {-\frac{\pi }{2},\frac{\pi }{2}} \right]$.  We set $\sigma _C^2 = \sigma _R^2 =-80$ dBm and the DFRC frame length is set as $L=256$. The power ratio matrix (PRM) for each user is a diagonal matrix, with each diagonal element following a distribution ${\cal C}{\cal N}\left( {0,{c_0}d_k^{ - \alpha }/L} \right)$, where ${c_0}=-40$ dB denotes the expected value of the average channel gain at the reference distance of 1 m, and $\alpha=2.8$ represents the path-loss exponent.

We consider three baseline schemes for comparison. (1) FPA: the antenna at BS and at each user are all fixed. (2) BS MA: the BS is equipped with $N_r$ MAs, while the position of antenna at each user remains fixed at the original point. (3) User MA: the BS is employed with FPA spacing by half of wavelength, while each user employ an MA. 
\begin{small}
\begin{figure}[htbp]
	\centering
	\begin{minipage}{0.85\linewidth}
		\centering
		\includegraphics[width=0.7\linewidth]{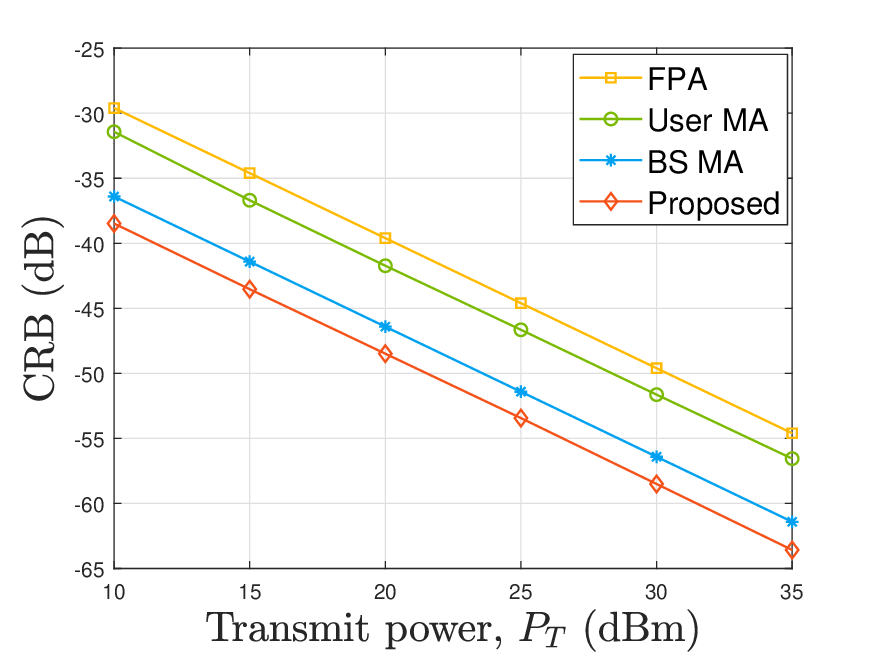}
		\caption{CRB versus the total transmit power $P_T$.}
	\end{minipage}
    \qquad
	\begin{minipage}{0.85\linewidth}
		\centering
		\includegraphics[width=0.7\linewidth]{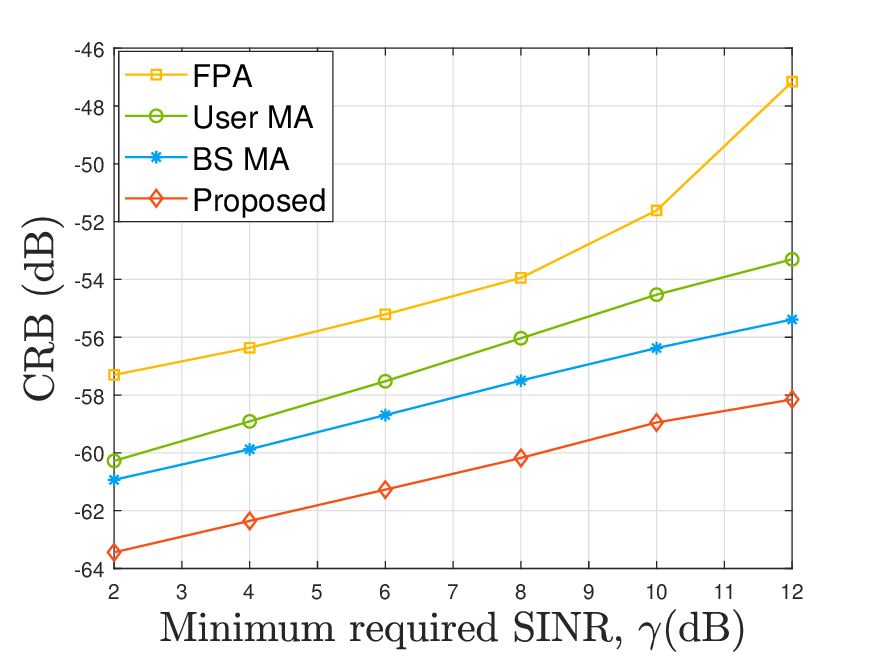}
		\caption{The CRB versus the required SINR.}
	\end{minipage}
    \qquad
	\begin{minipage}{0.85\linewidth}
		\centering
		\includegraphics[width=0.7\linewidth]{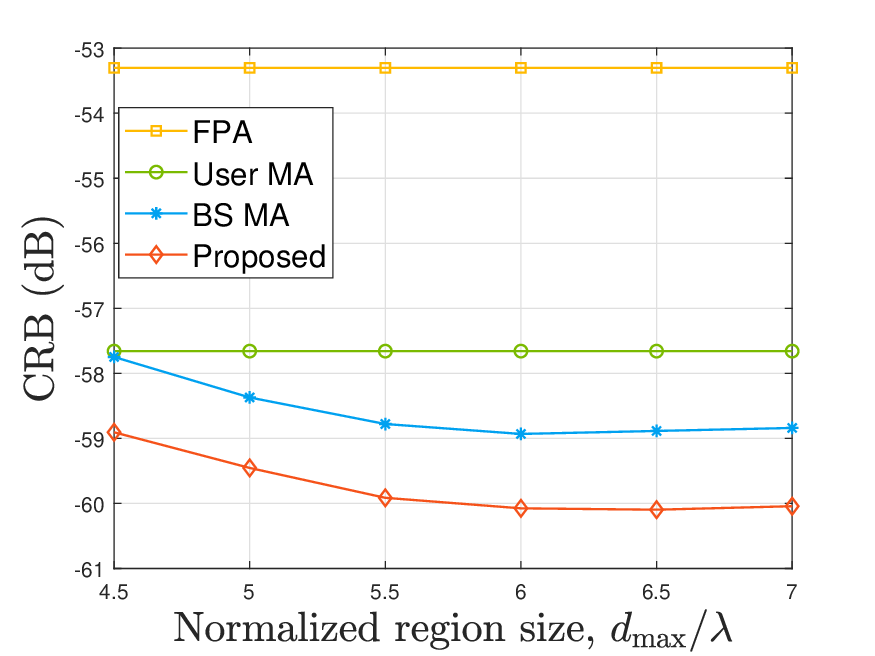}
		\caption{CRB versus the antenna moving region.}
	\end{minipage}
\end{figure}
\end{small}

Fig. 2 shows the CRB for target’s DoA estimation versus the BS transmit power when the SINR=10 dB. It can be seen that as the transmit power $P_0$ increases, the CRB in dB decreases monotonically in an approximately linear manner. This phenomenon can be attributed to the fact that a higher transmit power results in a greater beampattern gain in the target direction, which subsequently reduces the estimated error. It's also noteworthy that the deployment of MA can save more transmit power to achieve the same level of sensing accuracy compared with FPA.

Fig. 3 studies the the CRB versus the SINR threshold when the power budget is $P_T = 30$ dBm. It is observed that the CRB increases as the SINR increases which means there is a trade-off between communication performance and sensing performance under limited power conditions, where communication performance is improved at the expense of sensing performance. It is also observed that the deploymenr of MA can significantly reduce the CRB under the same SINR and the gap between the FPA and the three other benchmarks becomes larger when the SINR threshold is sufficiently large.
 
Finally, Fig. 4 describes the CRB variation versus the size of BS MA moving region. It is shown that the CRB slowly decreases as the moving region increases, which is due to the larger ${d_{\max }}$ creating a larger antenna aperture. Furthermore, the proposed scheme demonstrates superior performance in comparison to the BS MA scheme. This is due to the fact that users equipped with MAs are able to achieve a more robust channel condition, which allows the BS to allocate a reduced power level for communication user service and a greater power level for sensing.

\section{Concluision}
In this paper, we studied an MA-assisted multiuser ISAC system by exploiting the antenna position optimization and beamforming design. In particular, we derive the direct expression for the CRB w.r.t. the beamforming matrix and the BS antennas positions. Then, we formulated an optimization problem to minimize the CRB of target estimation under transmit power constraint, moving region constraints for MAs and SINR constraints for multiple communication users. To solve this non-convex problem, an AO algorithm combined with SDR approach and SCA technique was proposed to obtain a suboptimal solution by iteratively solving three subproblems. Simulation results validated that our proposed design with MAs can significantly reduce the CRB to improve sensing performance compared to that with the FPA arrays.

\ifCLASSOPTIONcaptionsoff
  \newpage
\fi

\bibliographystyle{IEEEtran}
\bibliography{name}
\end{document}